# THE BASIS FOR DESIGN OF A DSP-BASED COINCIDENCE SPECTROMETER

**Pham Dinh Khang[1], Nguyen Xuan Hai[2], Nguyen Nhi Dien[2], Pham Ngoc Tuan[2], Dang Lanh[2], Nguyen Duc Hoa[3], Nguyen An Son[3]**

[1] *Nuclear Training Center, 140 Nguyen Tuan, Hanoi*
[2] *Nuclear Research Institute, 01 Nguyen Tu Luc, Dalat*
[3] *University of Dalat, 01 Phu Dong Thien Vuong, Dalat*

**Abstract**
Coincidence technique and the coincidence measurement systems have been developed and applied for over 40 years. Most of popular coincidence measurement systems were based on analog electronics techniques such as time to amplitude conversion (TAC) or logic selecting coincidence unit. The above-mentioned systems are relatively cumbersome and complicated to use. With the strong growth of digital electronics techniques and computational science, the coincidence measurement systems will be constructed simpler but more efficient with the sake of application. This article presents the design principle and signal processing of a simple two-channel coincidence system by a new technique called Digital Signal Processing (DSP) using Field Programmable Gate Arrays (FPGA) devices at Nuclear Research Institute (NRI), Dalat.

## I. INTRODUCTION

The reduction of background of a nuclear measurement systems with active methods are based on coincidence or anti-coincidence techniques. These are mainly built from proper functional electronics modules in NIM or CAMAC standards [6, 8]. They allow us to identify coincidence or anti-coincidence events via main electronics blocks called coincidence unit or time to amplitude converter. Normally, basic configuration of a coincidence measurement system will, at least, consist of two channels, and the selection of coincidence event pairs depends on the defining moment of the pulses appearing at 'Timing' output of the pre-amplifier. Obvious drawback of this system is cumbersome in size, adjusting operation and synchronizing signals among electronics stages.

The growth of computer engineering and programmable devices that are capable of operating at high frequencies has allowed us to design a new spectrometry generation. The spectrometry generation is compact on size, simple in terms of connectivity and using [4, 7].

At Nuclear Research Institute (NRI), Dalat, a number of researches and construction of Compton suppression as well as event-event coincidence systems were performed in 1990s. The results of these studies were reported in ref. [3]. In the recent period at NRI, further studies on the coincidence spectrometer were presented in several publications [1, 2, 5]. Overall, although there have been significant improvements on acquiring as well as processing data, this system is still based on the traditional way in obtaining signals under the operation of a coincidence unit or TAC. With the research results gathered during the installation, investigation and exploitation of an 'event-event' coincidence spectrometer on digital signal processing (DSP) at NRI, we have now proposed a new approach for a design of the DSP-based multi-application coincidence technique through Field Programmable Gate Arrays (FPGA) devices. The basis for the design of this spectrometer will be presented within the framework of this article.

## II. THE BASIS AND METHOD FOR THE DESIGN

### 1. The basis for operation of an 'event-event' recording coincidence spectrometer using TAC

An 'event-event' recording coincidence spectrometer for processing data under a combination of the traditional analog electronics and interfacing unit is shown in Fig. 1. The standalone spectrometer is capable of linking to PC through PCI slot. The system consists of two channels: the upper channel and the lower one. All of the functional electronics modules are as follows:

 AMP 572A: Spectroscopy amplifier, model 572A, Canberra
 8k ADC 7072: Fast peak sensing ADC 8k, model 7072, Comp. Tec.
 HV 660: High voltage bias supply, model 660, Ortec
 TFA 474: Timing filter amplifier, model 474, Ortec
 CFD 584: Constant Fraction Discriminator, model 584, Ortec
 Delay 2058: Delay unit, model 2058, Canberra
 TAC 566: Time to amplitude converter, model 566, Ortec
 16k ADC 8713, Canberra: Peak sensing ADC 16k, model 8713, Canberra
 PCI7811R: Interfacing card, model PCI7811R.

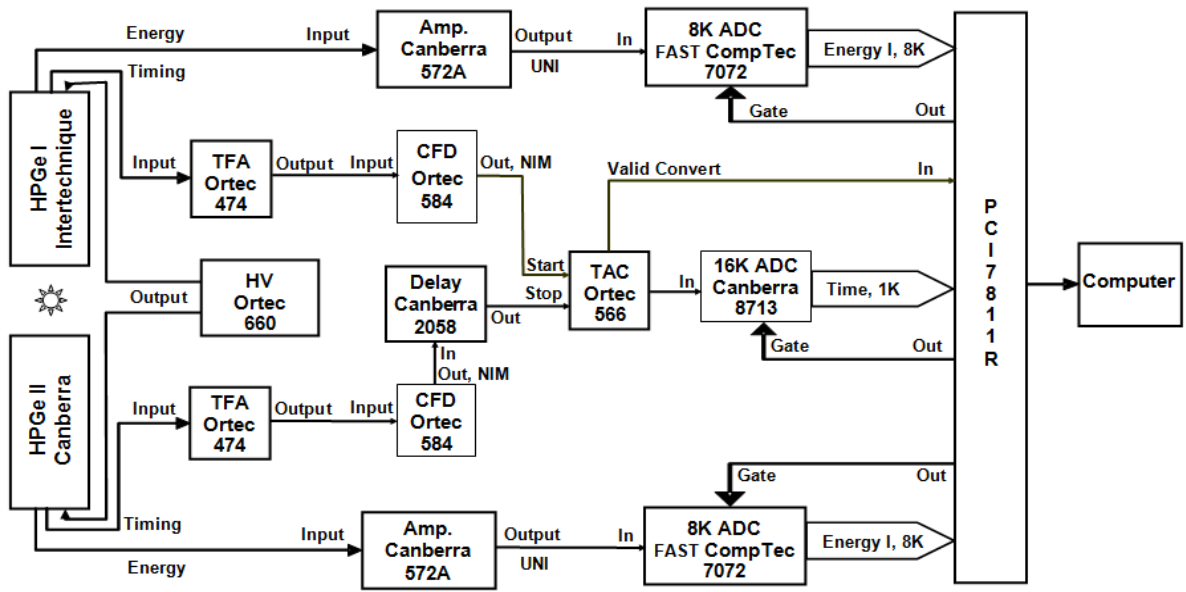

**Figure 1:** The block diagram of TAC-based coincidence spectrometer used at NRI [5].

The principal operation of the system is as follows (refer to Fig. 1): The signals appearing on the energy (E) outputs from two HPGe detectors are fed to both the inputs of two AMPs belonging to the upper and lower channels. In addition, two timing (T) outputs are also coming to the inputs of two TFAs. Next, the two-TFA outputs are fed to the inputs of CFDs. The upper CFD output will then strobe the 'Start' input TAC, and the other CFD output plays a role of stopping the TAC in one converting cycle for changing signal amplitude. When the interfacing unit 7811R receives 'Valid conversion' signal from TAC, it will soon send back three 'Gate' signals for gating all of ADCs; and at that time, ADCs are allowed to convert amplitude into BCD code numbers. After ADCs finish conversion cycles, the interfacing unit will read the code and write these values into memory.

After finishing write operation, ADCs return to an idle status and wait for another gate signal under the control of the next valid conversion signal. ADCs have no operation if there is no strobe signal at the gate regardless of incoming signals at their inputs. In a data file, the data will be arranged into three columns $E_1(n)$, $E_2(n)$ and $E_3(n)$. The values of $E_1(n)$ and $E_2(n)$ are in turn the amplitude code numbers of the two coincidence pulses coming from detector 1 and detector 2 respectively, $E_3(n)$ is a value corresponding to the differential time between two events, $n$ is the ordinal number of pairs of coincidence events from the beginning of measurement. After finishing the measurement, data is handled by the multivariable statistics processing program to obtain information about energy, transition intensity and decay scheme of nuclei on research.

Dead time (DT) of the system will be calculated as the shortest interval between two pairs of consecutive amplitude codes recorded via spectrometer. Total DT of the system depends on the sampling rate of ADC and data-transfer speed of interfacing unit. The slower they work, the longer dead time, and vice versa. Because DT is one of the causes affecting the system's efficiency, therefore it should be ensured that the shorter the DT the higher the data.

$$\tau_{min} = \tau_1 + \tau_2 + \tau_3 + \tau_4$$

where, $\tau_1$: delay of spectroscopy amplifier, $\tau_2$: shaping time, $\tau_3$: ADC's conversion time, and $\tau_4$: time for data transfer operation of the interfacing unit.

## 2. The principle for design of an 'event-event' recording digital coincidence spectrometer

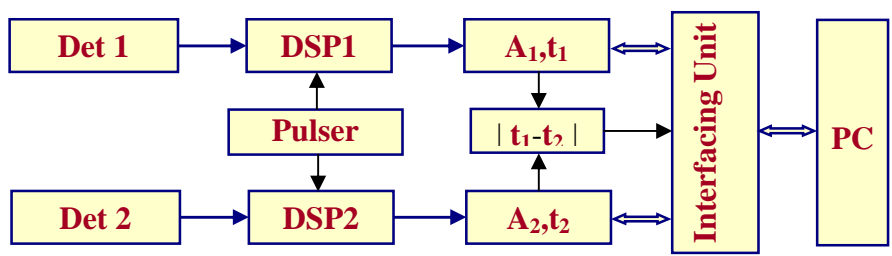

**Figure 2:** The block diagram of an 'event by event' recording digital coincidence spectrometer.

Fig. 2 shows the block diagram of an 'event-event' recording digital coincidence spectrometer. Its operation principle is as follows: when the radiation signals are recorded from detector 1 or detector 2, DSPs analyze the amplitude of the pulses and then give the corresponding values of ($A_1$, $A_2$). At the same time, when the signals exceed the lower thresholds, DSPs will read more the values corresponding to the moments of ($t_1$, $t_2$) at those the pulses exceed the aforementioned thresholds. The timing tester will determine the time difference between the two events $\Delta t = |t_1 - t_2|$. If $W$ is called *coincidence time window* of the system, there are a number of cases occurred as follows:

- $\Delta t \le W$: coincidence occurred. The programmer will write pair of events into memory with contents of *A1, A2* and $\Delta t$.
- $\Delta t > W$: no coincidence occurs. The program will remove pair of events.

In the second case, assuming that there was an event appeared beforehand on the first channel, the program will remove the values of *A1* and $\Delta t$ out of temporary memories and wait for the next event occurring on it. Value of $t_1$ will be compared to that of $t_2$ so as to determine the next pair of coincidence events. Thus, the process keeps on until the measurement finishes.

### III. ANALYSIS AND ASSESSMENT OF THE DESIGN

As physical side, test results for the block diagram in Fig. 1 are presented in Fig. 4. The test measurement was carried out with $^{60}$Co, activity ~ 20 kBq and $^{137}$Cs, activity ~ 100 kBq, measurement range of TAC is 500 ns. The obtained results showed that on the basis of the time difference of pairs of coincidence events between the two detectors, the spectra corresponding to $^{60}$Co and $^{137}$Cs can properly be separated with high precision. Drawback of the configuration is that the performance of the system reduced because of fixed time interval during the inputs of Start and Stop of TAC; therefore, a modification of circuit design which can replace the role of traditional start and stop signals had to be established. The modified circuit is called time-stamp one. In the second configuration, the flexibility of varying time stamp at which a triggering event occurred will allow us to overcome the aforementioned drawback. For the semiconductor detector when the incoming radiation interacts with the positions at the edge of crystal, the charge collecting time will tend increasing. This leads the rising edges of the pulses at the preamplifier's output pulled longer. If the operation frequency of the signal processing circuit is not high enough, or algorithm for determining threshold-crossing time is not so good (due to jitter or walk), then the timing resolution of the system will become worse significantly. If the frequency of the circuit is called $f_0$, error for determining time stamp will be $\pm 1/f_0$ (sec). For instance, if $f_0$ is equal to 40 MHz, the time error will be $\pm 25$ ns. To test this, PCI interfacing unit has been programmed for running at 80 MHz, the signals at the inputs (Start, Stop) of TAC were postponed with a number of different time intervals owning to nanosecond-delay module, Canberra. The final results showed that error in determining the threshold crossing time is $\pm 12.5$ ns.

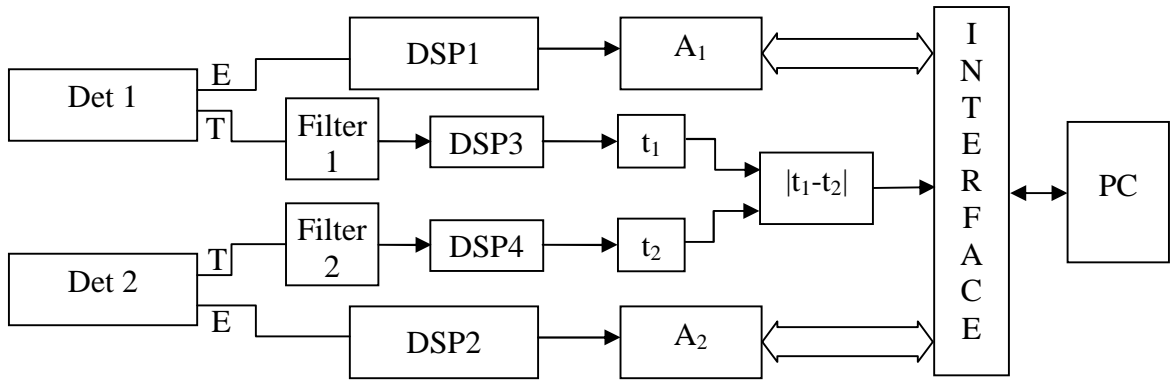

**Figure 3:** The principal design of digital 'event-event' coincidence system with addition of filtering time circuit.

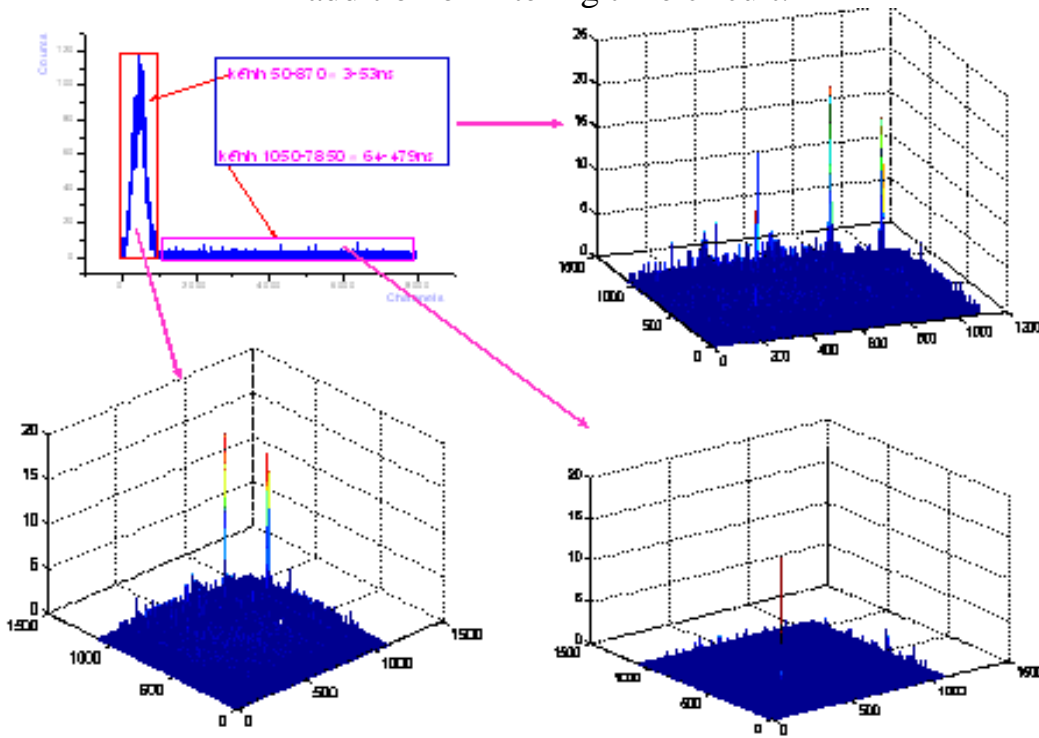

**Figure 4:** Test results as physical side for the configuration in Fig. 1.

To overcome the above-mentioned restrictions as using HPGe detectors, it is better to add an algorithm capable of removing the pulses that have the slow rising edges into DSPs. However, the additional algorithms can make processing speed to become slow, especially for determining the height of peak. Therefore, in this case the third configuration should be used in order to overcome this drawback (see Fig. 3). This design allows separating and processing properly timing signals. As a result, the pulses having the slow rising edges are removed via filtering circuits before coming to the digital processing parts will play a role in determining the threshold crossing time.

In case of having no coincidence signal, data reading algorithms have to be modified in order to remove amplitude value $A_1$ or $A_2$ because the filtering circuit removed almost of the non-suitable pulses.

## IV. CONCLUSIONS

The design of systems based on DSP techniques using FPGA allow to construct simple, compact and impact coincidence spectrometer in which all of parameters are selected and controlled by software.

Currently, the diagram in Fig. 2 has been studying and constructing at the Department of Nuclear Physics and Electronics, NRI, Dalat. It is hopeful that, in the near future, the design of system might be the basis for development and application of coincidence measurement techniques in the field of physics research and applications.